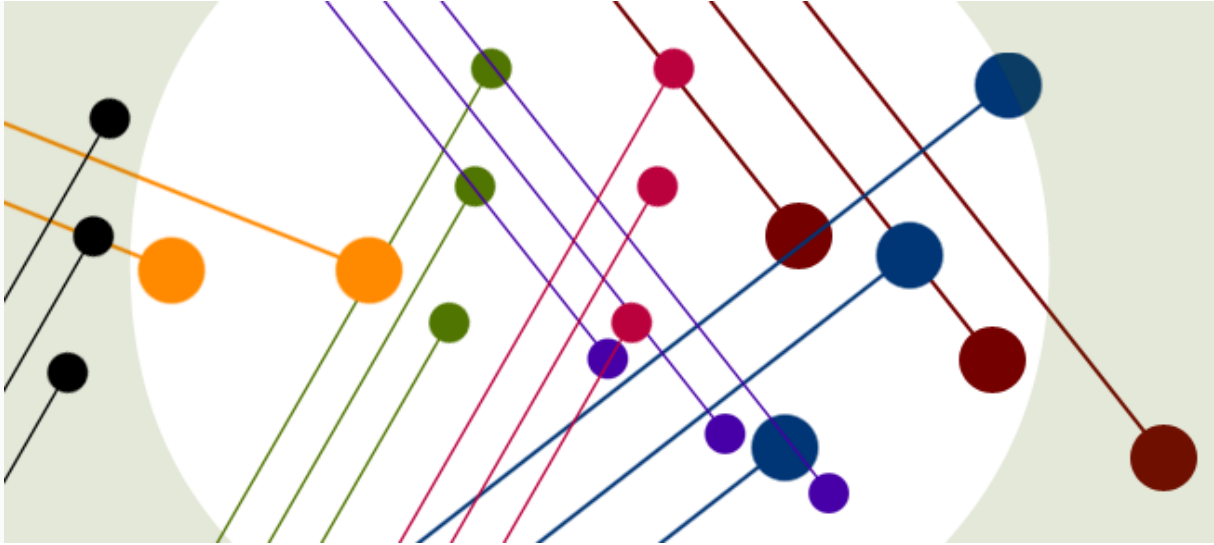

# Relationships in Large-Scale Graph Computing

*Article by Dan Petrovic*

Two years ago Grzegorz Czajkowski from Google's system infrastructure team published an article which didn't get much attention in the SEO community at the time. It was titled "Large-scale graph computing at Google" [1] and gave an excellent insight into the future of Google's search. Here I will highlight some of the little known facts which lead to transformation of Google's algorithm in the last two years.

## Everything is a Graph

Czajkowski argues that much of our existence can be represented in a graph form including social activities, personal relationship, professional activities, technology, transactions and other aspects of our lives, much like citations work among scientific papers. Google was the first company to successfully harness the power of the web by applying advanced graph analysis techniques including the concept of PageRank.

Here's an interesting fact. The concept of mapping society mathematically is nothing new. There are countless theoretical and even fictional references to calculations of socio-economic and behavioural elements in attempt to understand the humanity and map the future path of our civilisation. One such concept comes from the famous science fiction writer and a scientist Isaac Asimov in his Foundation series. Asimov's name for what Google's heading towards is "psychohistory".

> *"Psychohistory is a fictional science in Isaac Asimov's Foundation universe which combines history, sociology, and mathematical statistics to make general predictions about the future behavior of very large groups of people, such as the Galactic Empire. It was first introduced in the five short stories (1942–1944) which would later be collected as the 1951 novel Foundation." Source: http://en.wikipedia.org/wiki/Psychohistory_(fictional)*

In Asimov's fiction psychohistory is mapped towards a "prime radiant" which may deviate due to many variables and the scientists occasionally adjust the formula to factor in new variables.

## Dealing With Complexity

Static graphs in science can be extremely complex but when it comes to a dynamic environment such as society, knowledge, communication and information graphs exhibit perpetual growth in size and data/relationships which introduce a whole new dimension to the problem of scalability and computational power.

Google's solution to scalable graph analysis takes inspiration from the Bulk Synchronous Parallel Model in parallel processing. The framework is called Pregel and is capable of mining a wide range of graphs through a unique iterative vertex treatment. In Pregel each graph vertex works independently and can receive and send messages from and to other vertices at different stages of iteration which enables self-modification and mutation of graph's topology without having to re-run the entire process from scratch. With this framework Google is capable of incomparable computing scalability when it comes to graph data and it simplifies calculation of PageRank. Below is the code sample from the official Pregel paper [2]:

```
class PageRankVertex
    : public Vertex<double, void, double> {
 public:
  virtual void Compute(MessageIterator* msgs) {
    if (superstep() >= 1) {
      double sum = 0;
      for (; !msgs->Done(); msgs->Next())
        sum += msgs->Value();
      *MutableValue() =
          0.15 / NumVertices() + 0.85 * sum;
    }

    if (superstep() < 30) {
      const int64 n = GetOutEdgeIterator().size();
      SendMessageToAllNeighbors(GetValue() / n);
    } else {
      VoteToHalt();
    }
  }
};
```

Figure 4: PageRank implemented in Pregel.

Michael Nielsen [3] has recently covered Pregel and elaborated on its practical use (calculation of PageRank) and provides useful pseudo-code samples. According to Nielsen, one of the main benefits of the new framework is the fact that it removes the need for manual intervention from programmers as it is capable of scaling on a cluster autonomously. This leaves software engineers with more time to focus on the algorithm itself. This proves that Google hasn't changed their view on human element in their software and offers strong hint that even social factors will always remain just maths and nothing more.

## Computing PageRank

PageRank [4] is based on the random surfer model [5] where it is assumed that a person browsing the web follows links in a linear fashion until the point where their interest stops to the point where they stop browsing or abandon the current tree of research and start elsewhere (back to search results). With this logic (non-native) PageRank reduces with each click away from the source document. Naturally this is a simplified example in which there are no external links pointing to any of the documents along the browsing path. So for example in a website with a home page PageRank of 2 the next click may lead to a PageRank 1 page and finally to a zero value page at which point the "interest" to the user becomes minimal. Typical value of the PageRank dampening is 0.15. This tends to be more complex in a typical web document as there is usually more than one link on the page and PageRank ternds to circulate throughout the sites' navigation (sitewide links).

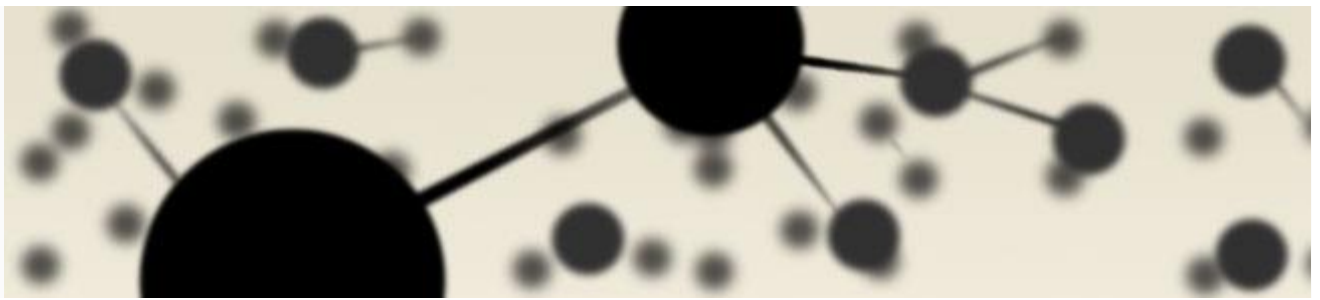

In Pregel aided PageRank calculation main data input is the web graph generated through crawler based activities. In practical terms, one can treat the web as "the graph", all of its documents (pages and indexable files) as "vertices" and links as "edges". In this model vertices (web documents) are typically initialised with a starting value. The interesting point is that this initial value makes no significant impact on the end-result in the computational model. After initialisation Pregel runs through supersteps by updating own value and sending messages to other vertices.

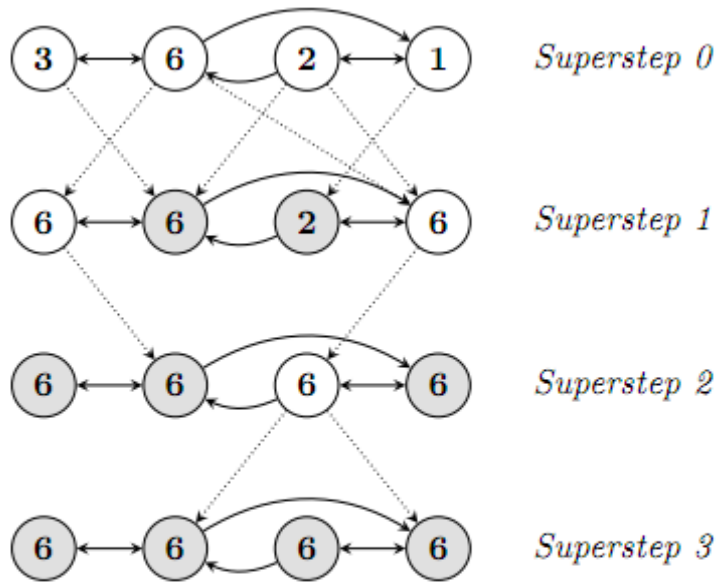

Figure 2: Maximum Value Example. Dotted lines are messages. Shaded vertices have voted to halt.

Nielsen also points out that each vertex also has a halting property, a value which determines the status as active or inactive (at which point the computation stops).

## Related Frameworks & Methodologies

Bill Slawski from SEO by the Sea commented on this article and submitted a presentation by David Konerding from Google titled: "Think Like a {Vertex, Column, Parallel Collection}"[8]. This document proves that there is much more behind Google and in addition to Pregel (the large-scale graph processing framework) they employ interactive analysis of web-scale datasets through Dremel and utilise FlumeJava for data-parallel pipelines.
Due to limitations in flexibility and application of MapReduce [7], software engineers had to design their own toolkits and frameworks for data-intensive parallel processing, typically with multi-step graph operations on a large scale and complex flow datapipes. It was interesting to find out that they also had to employ own tools to assist data analysts who deal with enormously complex (trillion-row) datasets.
The Pregel part of the presentation gives a good overview and introduces its similarity with Parallel Boost Graph [6]. From what we understand Google uses Pregel widely within their systems as it's easy to program and 'expressive'.
Benefits:

- Breadth-first search
- Strongly connected components
- PageRank Compatibility
- Label propagation algorithms
- Minimum spanning tree
- Δ-stepping parallelization of Dijkstra's SSSP algorithm

- Several kinds of vertex clustering
- Maximum and maximal weight bipartite matching

# Dremel: Think Like a Column

There is no better way to explain the concept other than by illustration. In the 3D renderings below you can see the logic behind the record-oriented and column-oriented approach:

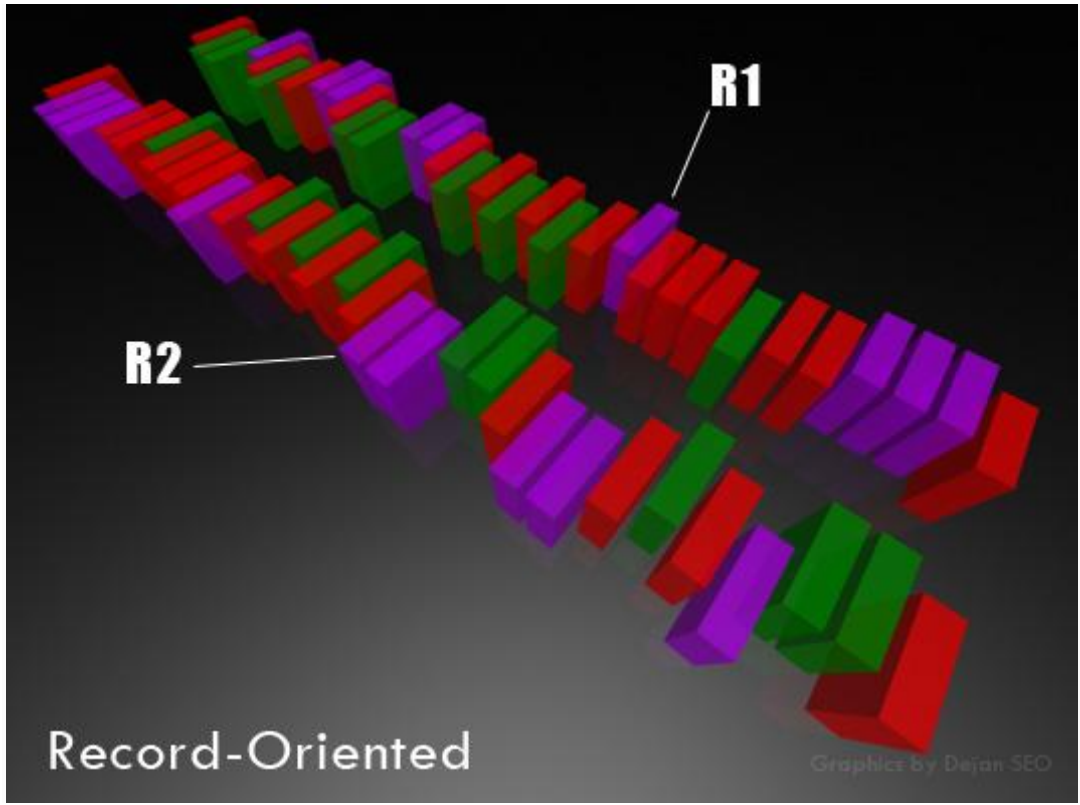

In the next example recordsets are arranged in distinct columns:

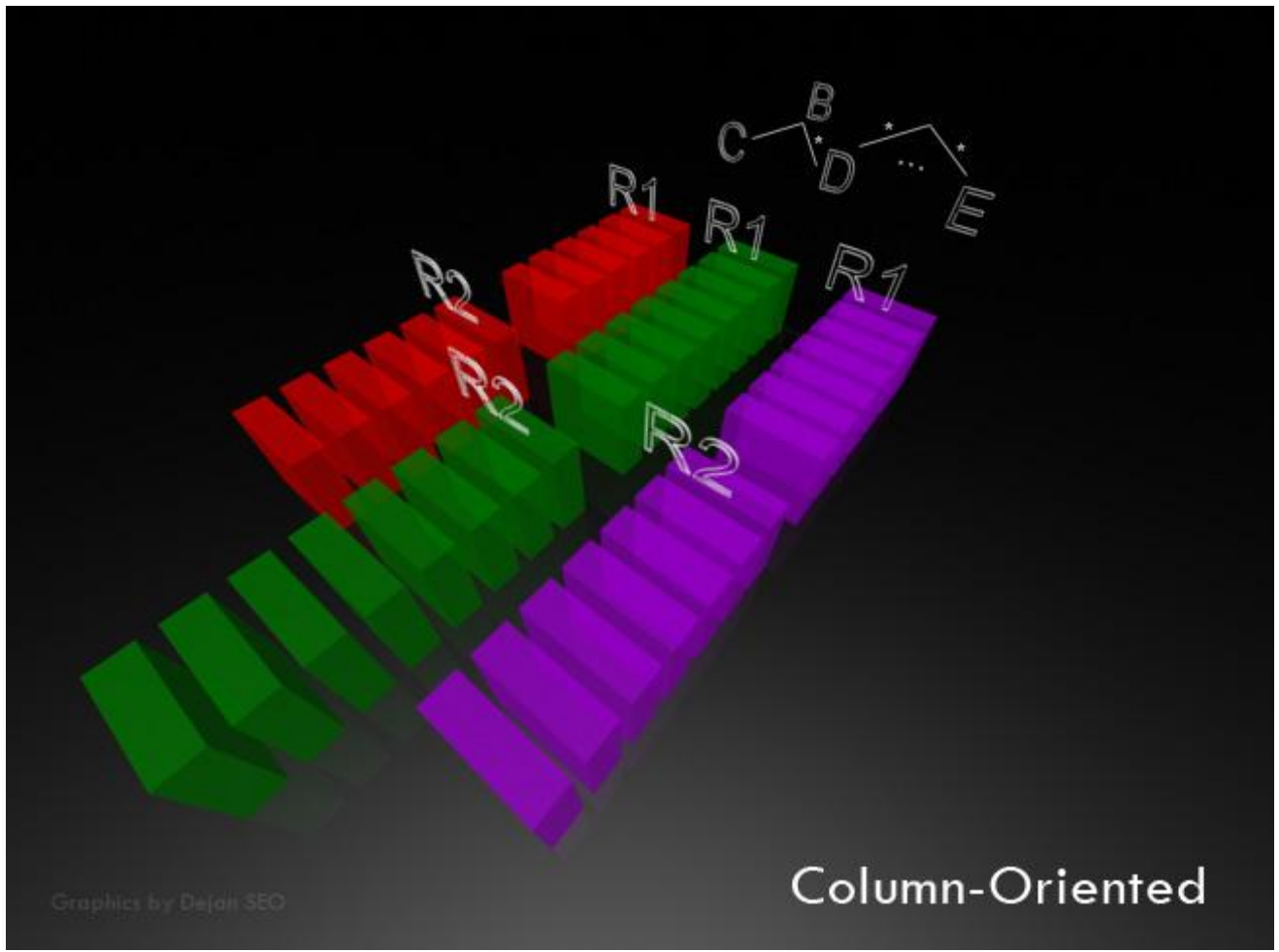

Characteristics and Benefits of Dremel:

- Trillion-record, multi-terabyte datasets
- Scales to thousands of nodes
- Interactive speed
- Nested data
- Columnar storage and processing
- In situ data access (e.g., GFS, Bigtable)
- Aggregation tree architecture
- Interoperability with Google's data management tools (e.g., MapReduce)

Practical Application of Dremel at Google:

- Analysis of crawled web documents
- Tracking install data for applications on Android Market
- Crash reporting for Google products
- OCR results from Google Books
- Spam analysis
- Debugging of map tiles on Google Maps
- Tablet migrations in managed Bigtable instances
- Results of tests run on Google's distributed build system
- Disk I/O statistics for hundreds of thousands of disks
- Resource monitoring for jobs run in Google's data centers
- Symbols and dependencies in Google's codebase

## FlumeJava: Think Like a Parallel Collection

FlumeJava was released to Google users in May 2009 and now hundreds of pipelines who process gigabytes to petabytes are run by hundreds of users every month. It is easier to use than MapReduce and can control optimizer and executor when needed (as well as better handling of unpredicted situations). Example below is for TopWords:

```
readTextFile("/gfs/corpus/*.txt")
    .parallelDo(new ExtractWordsFn())
    .count()
    .top(new OrderCountsFn(), 1000)
    .parallelDo(new FormatCountFn())
    .writeToTextFile("cnts.txt");

FlumeJava.run();
```

Qualities Google is after are fault tolerance by design and handling the processes in such way that if individual node fails it merely slows down completion (instead of stopping it altogether). Everything is large-scale by design and architecture (trillions of rows, billions of vertices, petabytes of data). Tools and systems within Google are interchangeable and used by multiple groups.

## Conclusion

There is no doubt that this is only a glimpse at Google's true complexity in their efforts to organise the world's information. Frameworks of this kind enable Google to maximise the use of their resources and provide superior results in comparison to their competitors, not just due to crude computing and processing power but also thanks to intelligent software solutions. Google is still not abandoning search as their first priority and their next obvious target with this great framework is to understand social interactions online – and in the way that nobody else can.

For Google, all the stars are aligned, they have the technology, money and all the data in the world. Soon their technology will likely reach the level incomprehensible to us ordinary humans. Credit goes to two technological visionaries Ray Kurzweil (Technological Singularity) and Isaac Asimov (Psychohistory) who predicted it all years ahead of us all, and I shall end with my favorite quote:

**"Any sufficiently advanced technology is indistinguishable from magic.", Arthur C. Clarke**